\documentclass[nofootinbib,floats,aps,preprint,secnumarabic]{revtex4}
\usepackage{graphicx,amssymb}

\newcommand{\intz}{\int_{}^{}dz}
\newcommand{\ben}{\begin{enumerate}}
\newcommand{\een}{\end{enumerate}}

\newcommand{\pd}[2]{\frac{\partial #1}{\partial #2}}

\newcommand{\z}{z}

\newcommand{\bea}{\begin{eqnarray}}
\newcommand{\eea}{\end{eqnarray}}
\newcommand{\beqa}{\begin{eqnarray}}
\newcommand{\eeqa}{\end{eqnarray}}
\newcommand{\beq}{\begin{equation}}
\newcommand{\eeq}{\end{equation}}
\newcommand{\bay}{\begin{array}}
\newcommand{\eay}{\end{array}}

\def\gsim{\ \rlap{\raise 3pt \hbox{$>$}}{\lower 3pt \hbox{$\sim$}}\ }
\def\lsim{\ \rlap{\raise 3pt \hbox{$<$}}{\lower 3pt \hbox{$\sim$}}\ }

\def\parsla{\partial\!\!\!\slash}

\def\le{\left}
\def\ri{\right}
\def\half{\frac{1}{2}}
\def\quart{\frac{1}{4}}

\def\dmu{\partial_{\mu}}
\def\d5{\partial_5}

\begin{document}

\preprint{\vbox 
{\hbox{} 
\hbox{} 
\hbox{} 
\hbox{hep-ph/0507036(v2)}}}

\vspace*{3cm}

\title{Higgs Localization in Split Fermion Models}

\def\addtech{Physics Department,
Technion -- Israel Institute of Technology,\\
Haifa 32000, Israel\vspace*{2cm}}

\author{\vspace*{1cm} Ze'ev Surujon}\affiliation{\addtech}

\begin{abstract}
The flavor puzzle of the Standard Model is explained in split fermion
models by having the fermions localized and separated in an extra
dimension.  Many of these models assume a certain profile for the
Higgs VEV, usually uniform, or confined to a brane, without providing
a dynamical realization for it. By studying the effect of the
coupling between the Higgs and the localizer fields, we obtain these
scenarios as results, rather than ans\"{a}tze.  Moreover, we discuss
other profiles and show that they are phenomenologically viable.
\end{abstract}
\maketitle


\section{Introduction}
\setcounter{equation}{0}
One of the indications for the incompleteness of the Standard Model (SM) is
the hierarchy among its flavor parameters.
An attractive solution for this flavor puzzle is presented in split fermion
models~\cite{AS}, where the fermion zero modes are split over
an extra dimension.
The effective 4D Yukawa couplings are then suppressed by exponentially small
overlaps between wavefunctions of different fermion zero modes.
In some of these models~\cite{MS,ASmodels,GP} the fermions
are localized at various positions in $x_5$, while in others~\cite{KT},
they are attached to 4D planes, with an exponential
penetration into the bulk.
A key ingredient in all these models is a real scalar field (the localizer)
which acquires a bulk dependent VEV.
More examples, discussions and experimental signatures of split fermion models
can be found in~\cite{PDG,GGH,GaussLocH,LocH,AGS}.

The Higgs VEV is usually assumed either uniform~\cite{MS,GP,GGH,KT}, or
confined to a brane~\cite{KT,GaussLocH,LocH}.
While in some works~\cite{DS,KT} there are ideas about generating its profile,
no comprehensive study was done.
In order to motivate a certain profile, one has to study the coupling between
the Higgs and the localizer, which is the purpose of this work.
We work at the classical level, where we were able to realize the above two
scenarios.
A uniform Higgs is obtained if the above coupling is small, which requires
fine tuning of the model parameters.
Conversely, a scenario in which the Higgs is localized at a brane is
obtained without fine tuning.
We discuss an explicit solution of this sort in the large orbifold limit,
and its phenomenological constraints.
Finally, we show that there are other solutions which are not fined tuned and
are phenomenologically viable.

\section{Split Fermions}

\subsection{Overview}
Our interest lies in split fermion models with one extra dimension.
While there is an extensive typology of such models, a feature common to
many of them is the appearance of a real scalar `$\Phi$' called the localizer,
which is a SM singlet.
The remaining field content is similar to that of the SM, but with Dirac
fermions (there are no chiral representations in five dimensions).
By coupling to the fermions, the localizer VEV serves as a position dependent
mass which vanishes at the 3-brane where the fermion is localized.
In order to see that, we write the relevant part of the Lagrangian as
\beq
   {\cal L}=\bar \Psi_i\le(i\delta_{ij}\Gamma^M\partial_M-
   \frac{\lambda_{ij}}{\sqrt{M_*}}\langle\Phi\rangle-M_{ij}\ri)\Psi_j,
\eeq
where $\Gamma^M$ are the five Dirac matrices, $M_*$ is the UV cutoff so that
the $\lambda_{ij}$ are dimensionless, and $i,j$ are flavor indices.
As a first step we discuss an infinite extra dimension, where we denote the
extra dimension coordinate as `$z$'.
A further simplification is to take $[\lambda,M]=0$.
Then we can work in the mass basis where both $\lambda$ and $M$ are diagonal
(this assumption is relaxed in~\cite{twist}).

We need $\Phi(z)$ to be a topological defect with co-dimension
$D-4$ in order to confine the other fields to a 4D universe.
In an infinite extra dimension we assume the kink solution,
\beq
   \Phi(z)=\frac{\mu}{\sqrt{\lambda_{\phi}/M_*}}
   \tanh\le(\frac{\mu z}{\sqrt{2}}\ri),
\eeq
where $\mu$ is the 5D mass parameter of the localizer and $\lambda_{\phi}$ is
its dimensionless quartic coupling.
The equations of motion (EOM) for the Kaluza-Klein (KK) wavefunctions of the
fermions are obtained by solving the 5D EOM with separation of variables.
In particular, the resulting wavefunctions for the zero modes are
\beq
   f^{i(0)}_{L,R}(z)=N_{L,R}\exp\left(\mp\int_0^z \le[\lambda_i\Phi(z')+M_i\ri]
   \, dz'\right),
   \label{eq:wf-generic}
\eeq
where $M_i(\lambda_i)$ are the eigenvalues of $M(\lambda)$.
We can see that only one of each wavefunction pair is normalizable, depending
on the sign of $\lambda_i$.
In order to get rid of the mirror fermions in a realistic model,
a common solution is to compactify the extra dimension on a $S^1/Z_2$ chiral
orbifold~\cite{GGH,KT,GP}.
The orbifold boundary conditions force right handed fermions to be odd,
which is incompatible with Eq.(\ref{eq:wf-generic}), and therefore
the right handed zero mode is projected out of the spectrum.

Having localized the SM chiral fields at various points in $x_5$, we now turn
to show how small 4D Yukawa couplings arise naturally in this setup.
For example, in the Arkani-Hamed--Schmaltz (AS) model, the relevant part
of the Lagrangian for the leptons is
\beqa
   {\cal L}_{\rm 5D} &=&
   \bar L_i\le(i\parsla_5-M_L- \frac{\lambda}{\sqrt{M_*}}\Phi\ri)L_i
    + \bar E_i\le(i\parsla_5-M_E+\frac{\lambda}{\sqrt{M_*}}\Phi\ri)E_i
    \nonumber \\ &-& \le(\frac{Y_{ij}}{\sqrt{M_*}} \bar L_i H E_j+c.c.\ri),
    \label{eq:AS_lag}
\eeqa
where $Y_{ij}$ is the 5D Yukawa couplings with the Higgs.
Upon KK decomposition, we obtain the effective low-energy 4D Lagrangian,
\beq
   {\cal L}_{\rm 4D} =
   \le(\frac{Y_{ij}}{\sqrt{M_*}}\int dz\ f^{\ell*}_i(z)
   \le[f^h(z)\ h(x)+v_H(z)\ri]f^e_j(z)\ri)\bar\ell_L^i(x)\ e_R^j(x) + c.c.\;,
\eeq
where the zero modes are denoted in small Latin letters, and $f^i(z)$
are their wavefunctions.
Note that in the KK decomposition of the Higgs field we distinguish between
the VEV and the wavefunction of the lowest mode, which we call the {\it
ground} KK mode.

As already mentioned, in realistic models the extra dimension is compactified
on a $S^1/Z_2$ chiral orbifold, where the localizer and the right handed
fermions are odd under the $Z_2$ reflection.  The Higgs and the left handed
fermions are even.
Since 5D Dirac masses are forbidden by these boundary conditions, the fermions
cannot be split up in the usual way, but a scenario with fermions in the bulk
can still be constructed~\cite{KT,GP}.
Another approach is presented in~\cite{KT}, where the localizer VEV has a
narrow domain wall, effectively a step function, and the Higgs VEV is confined
to one of the fixed points.
The fermion wavefunctions in this model are localized at each of the fixed
points, with an exponential decay toward the other fixed point.
The sign of the coupling to the localizer determines the localization point,
and its magnitude fixes their width.
More specifically, in this model the third generation quark doublet and the
top singlet are localized at the Higgs brane whereas the other quarks are
localized at the other fixed point.
The 4D effective Yukawa is then proportional to the value of the wavefunction
at the Higgs brane, thus giving a large Yukawa only for the top Yukawa,
since both the top and $Q_3$ are maximal at the Higgs brane.
In the next section we obtain such scenario as a specific classical solution.

In a milder, ``regularized'' version of the above, the Higgs VEV is rather
localized, and not a delta function.
A non-uniform Higgs VEV is possible as long as the mass of the $W^{\pm}$
ground mode is predicted correctly, namely,
\beq
   g_4^2v^2_{\rm EW} = \frac{g_5^2}{M_*}
   \int dz\, \le|h(z)f_W(z)\ri|^2, \label{eq:phen}
\eeq
where $g_D$ is the $D$ dimensional SU(2) gauge coupling, $h(z)$ is the 5D
Higgs VEV, and $f_W(z)$ is the wavefunction of the $W^{\pm}$ ground mode.
The relation between the 4D and 5D gauge couplings is given by
\beq
   g_4=\frac{g_5}{\sqrt{M_*}}\int_0^Ldz\, f_W(z)\le|f_{\psi}(z)\ri|^2
   \simeq \frac{g_5}{\sqrt{M_*L}},
\eeq
provided that the $W^{\pm}$ wavefunction is nearly flat.
Therefore the gauge couplings drop out and the phenomenological constraint
Eq.(\ref{eq:phen}) reads
\beq
   v^2_{\rm EW}=\int_0^L dz\, h^2(z). \label{eq:phen1}
\eeq
This condition is necessary in order to comply with phenomenology.

We treat our model as an effective field theory, and therefore the presence
of nonrenormalizable terms does not pose an essential problem.
Throughout this work we consider operators which translate into renormalizable
terms in the effective four dimensional theory, such as
$\phi\bar\psi\psi$ (dimension $5\half$ in the 5D theory), or
$\phi^4$ (dimension 6 in the 5D theory).
Higher dimension operators are suppressed by appropriate powers of $p/M_*$
and therefore can be neglected in the low energy limit, where the theory is
effectively four dimensional.

Note that here we only give a classical treatment of split fermions,
following the literature (see for example \cite{AS,MS,KT}).  We do not
discuss quantum corrections, and in particular anomalies in extra
dimensions. Discussions concerning such issues can be found, for example,
in~\cite{anom}.

\subsection{Tree Level FCNC} \label{FCNC}
In SM extensions there can be various sources for tree-level
Flavor Changing Neutral Currents (FCNCs).
Such interactions can be mediated by the $Z^0$, by the Higgs or by new
bosons.
In split fermion models, every KK mode of the neutral gauge bosons or the
Higgs can mediate FCNC.
In particular, the KK modes of the gauge fields have $x_5$-dependent
wavefunctions and therefore their couplings to the fermion zero modes are
flavor dependent.
The resulting FCNCs provide a bound on the size of the extra
dimension~\cite{KT}.
Here we are concerned with Higgs mediated FCNCs.
In general, a scalar can mediate FCNC if its Yukawa term is not aligned with
the fermion mass term.
As an example we can think of the Higgs fields in multi Higgs models.
In split fermion models the case is similar, although there is only one Higgs.
This can be explained from a 5D or from a 4D point of view.

From the 5D point of view, the Higgs field is expanded about its VEV as
\beq
   H(x,z)=v_H(z)+\tilde H(x,z).
\eeq
The shifted field $\tilde H$ has vanishing VEV, and we decompose it into
KK modes,
\beq
   \tilde H(x,z)=\sum_n h_n(x)f_n(z).
\eeq
Retaining only the ground KK mode ($n=0$) we see that while the effective 4D
Yukawa couplings are given by
\beq
   y_{ij}=Y_{ij}\intz\ \frac{f^h(z)f^{i*}(z)f^j(z)}{\sqrt{M_*}},
\eeq
the fermion mass term induced by the Higgs VEV is
\beq
   m_{ij}=Y_{ij}\intz\ \frac{v(z)f^{i*}(z)f^j(z)}{\sqrt{M_*}},
\eeq
leading to $y_{ij} \not\propto m_{ij}$.
This misalignment between the 4D Yukawa and mass matrices implies FCNC at
the Lagrangian level.

In order to see the above from a 4D point of view, we distribute the non
uniform Higgs VEV among the different KK modes of the Higgs.
That is, we write
\beq
   H(x,z)=\sum_n \le[v_n+h_n(x)\ri]f_n(z)
\eeq
with
\beq
  v_H(z)\equiv \sum_n v_n f_n(z).
\eeq
Such 4D theory contains many KK fields~($h_n$), each with its own VEV~($v_n$).
The 4D mass term is given by
\beq
   m_{ij}=Y_{ij}\sum_n \frac{v_n}{\sqrt{M_*}}\intz\ f^h_n(z)f^{i*}(z)f^j(z),
\eeq
while the Yukawa coupling to the $n$-th Higgs KK is
\beq
   y^n_{ij}=\frac{Y_{ij}}{\sqrt{M_*}}\intz\ f^h_n(z)f^{i*}(z)f^j(z).
\eeq
The resulting Lagrangian is that of a multi Higgs model, but without
natural flavor conservation to prevent FCNC.


\section{The Scalar Sector}
Split fermions, then, provide a mechanism to localize the fermion zero modes.
A side effect, however, is that a similar mechanism can (and therefore does)
apply for the SM Higgs, as the latter couples to the localizer and inevitably
gets localized.
In order to find classical solutions for the Higgs VEV, we should
study the scalar sector, which includes the SM Higgs and the localizer.
More specifically, we are interested in the case of two scalars in one
spatial dimension.
In this work, for simplicity we replace the SM Higgs (four degrees of freedom
before electroweak symmetry breaking) with a real field
(one degree of freedom).
We start with the potential
\beq
   U(\phi,h)=-\half \mu^2\phi^2+\quart\lambda\phi^4
   -\half\mu_h^2 h^2 + \quart\lambda_h h^4
   +\half g\phi^2 h^2.   \label{eq:two_pot}
\eeq
The application to the five dimensional model involves
adding the appropriate powers of the 5D cutoff scale $M_*$.
For example, $\lambda,\lambda_h,g$ are couplings with mass dimension $(-1)$.
For now we work in natural units where $M_*=1$.

Starting with the simplistic AS model, where the extra dimension is infinite,
we know that the $g\to 0$ limit leads to the uniform Higgs solution,
\beq
   \phi(z)=\frac{\mu}{\sqrt{\lambda}}\tanh \le(\frac{\mu z}
   {\sqrt{2}}\ri)   ; \qquad  h(z) \equiv \frac{\mu_h}{\sqrt{\lambda_h}}.
\eeq
We also keep in mind that we seek solutions in which
the localizer $\phi(z)$ is antisymmetric and the Higgs $h(z)$ is symmetric
in order to match the orbifold boundary conditions upon compactification.

\subsection{A Uniform Higgs}

Split fermion models require that there is a hierarchy between the
localizer and the Higgs scales, in order that the fermion KK modes
would not acquire ${\cal O}(m_{\rm EW})$ masses.
The bound on the $v_{\rm EW}/\mu$ ratio depends on the details of the model,
but characteristic values are roughly
$v_{\rm EW}/\mu\sim (0.1\;{\rm TeV}/100\;{\rm TeV})$~\cite{KT}.
Note that this is a direct bound on $\mu$, unlike the bound related to
gauge KK modes, which constrains the size of the extra dimension.
The $v_{\rm EW}\ll\mu$ hierarchy suffers from fine tuning, since
both scales get radiative corrections proportional to the UV cutoff.
We also mention that the observation of universality in the weak interactions
puts a bound on $m_{EW}/\mu$ which is of similar order.

\begin{figure}[t]
  \centering
  \includegraphics{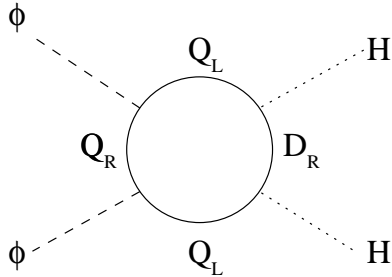}
  \caption{\small A one-loop contribution to $\phi^2H^{\dagger}H$.}
  \label{fig:box}
\end{figure}

In uniform Higgs scenarios, there is another fine tuning problem, related to
the generating of the coupling $g$ in $g\phi^2H^{\dagger}H$.
This operator is corrected even in the $g_{\rm tree}\to 0$ limit.
In one loop the only diagrams which contribute are those with fermions
running in the loop (see Fig.~\ref{fig:box}).
Since such diagrams depend strongly on the UV cutoff, the coupling
$g_{\rm tree}$ must be fine tuned in order to cancel the radiative corrections.
Note that models with uniform Higgs require the above fine tuning
{\it in addition} to the $\mu_h\ll\mu$ related fine tuning,
which must be assumed in any split fermion model.
In the rest of this work we do not consider quantum corrections, namely,
we work exclusively at the classical level.

\subsection{A perturbative/adiabatic approximation}

\begin{figure}[t]
   \centering
   \includegraphics[width=0.6\textwidth]{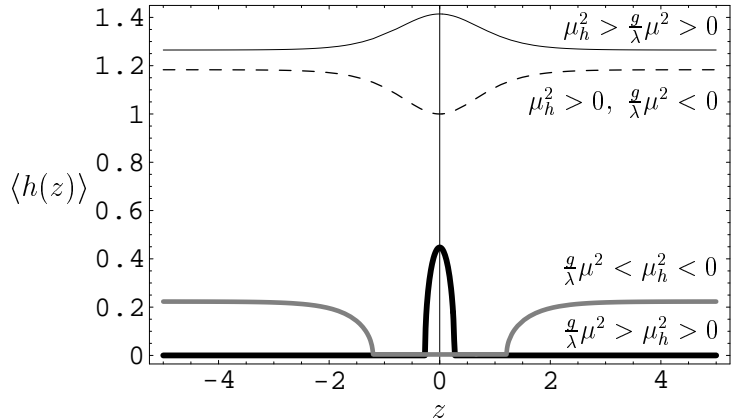}
   \caption{\small Four scenarios according to the perturbative approximation.
   The Higgs VEV is illustrated for four different regions in the parameter
   space of the scalar potential.
   As can be seen, the Higgs VEV can be attracted to or repulsed from
   the core of the domain wall.  The graphs are in arbitrary units.}
   \label{fig:adiabatic}
\end{figure}

Going back to the classical problem, the equations of motion,
\beq
   \phi''=\pd{U(\phi,h)}{\phi} ; \qquad h''=\pd{U(\phi,h)}{h},
   \label{eq:eom2}
\eeq
are coupled and cannot be integrated in a straightforward manner.
However, a particularly simple scenario is obtained when $gh(z)^2\ll\mu^2$,
that is, when the Higgs is affected by the localizer but not vice versa.
This limit is realized by the condition
\beq
   \frac{g}{\lambda_h}\mu_h^2\ll\mu^2. \label{eq:region}
\eeq
Then we can approximate a solution as follows:
The localizer ($\phi$) is assumed to be the kink,
\beq
   \phi(z)=\frac{\mu}{\sqrt{\lambda}}\tanh \le(\frac{\mu z}
   {\sqrt{2}}\ri),
\eeq
and we wish to find the Higgs VEV.
The Higgs potential is then given by
\beq
   U(h)=\half M^2(z)h^2 +\frac{\lambda_h}{4}h^4,
\eeq
where
\beq
   M^2(z) \equiv -\mu_h^2+\frac{g\mu^2}{\lambda}
      \tanh^2\le(\frac{\mu z}{\sqrt{2}}\ri)  \label{eq:mass}
\eeq
stands for the squared bulk mass of the Higgs.
The equation of motion for the static Higgs VEV is then
\beq
   h''=\lambda_h h^3+M^2(z)h.  \label{eq:higgs_eom}
\eeq
We expect the solution for $h(z)$ to get perturbative corrections of order
${\cal O}\le(v_H/v_\phi\ri)$.
Unfortunately the zeroth order is already hard to solve.
A further approximation is to neglect the left hand side of
Eq.(\ref{eq:higgs_eom}).
In this ``adiabatic'' approximation, only in the regions
where $M^2(z)<0$, the Higgs develops a VEV which is simply
\beq
   h(z)=\sqrt{\frac{-M^2(z)}{\lambda_h}}
   \le[1+{\cal O}\le(\frac{v_H}{v_\phi}\ri)
   +{\cal O}\le(\frac{h''(z)}{\mu_h^2}\ri)\ri].
\eeq
Note that in this approximation $h''(z)$ diverges where $M^2(z)=0$.
At these regions we expect large deviations.  At other regions, $h''(z)$
is not very large.

With the above result, in which we neglected the curvature,
we recognize four scenarios, depending on the parameters
(See Fig.~\ref{fig:adiabatic}).
Among these scenarios, we can identify one where the Higgs is localized
at the domain wall.
An exact solution of this form is obtained below, using a mechanical
analogy~\cite{coleman,rajaraman}.
Before going on to the mechanical analogy for two fields, we recall the
simpler case of one real scalar.

\subsection{One Scalar field} \label{localizer}
\begin{figure}[t]
    \centering
    \includegraphics{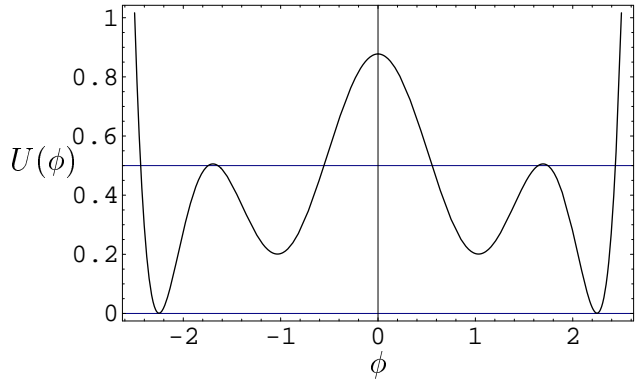}
    \caption{\small A potential with local minima located between two
             global minima.}
    \label{fig:orbit-1d}
\end{figure}

\begin{figure}[t]
   \centering
   \parbox{0.51\textwidth}{\raisebox{4mm}
      {\includegraphics[width=0.51\textwidth]{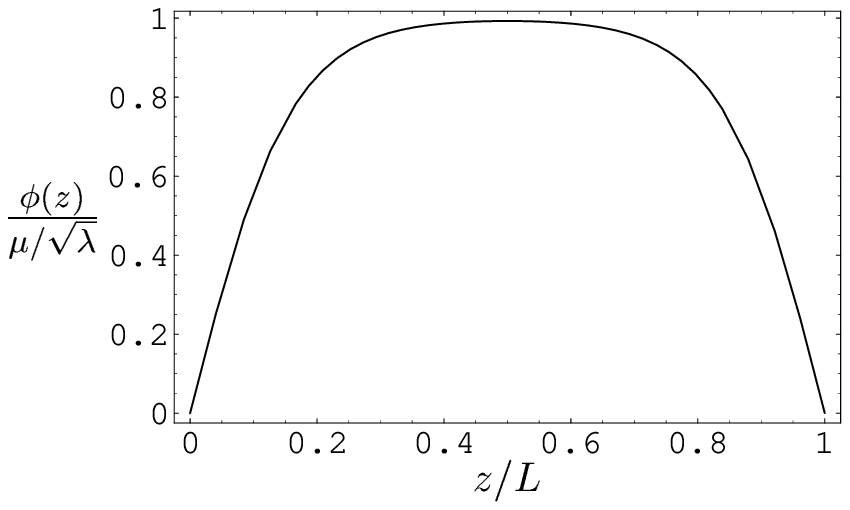}}
   \centerline{\hspace{15mm}\bf (\ref{fig:kak}a)}}
   \parbox{0.48\textwidth}{\includegraphics[width=0.48\textwidth]{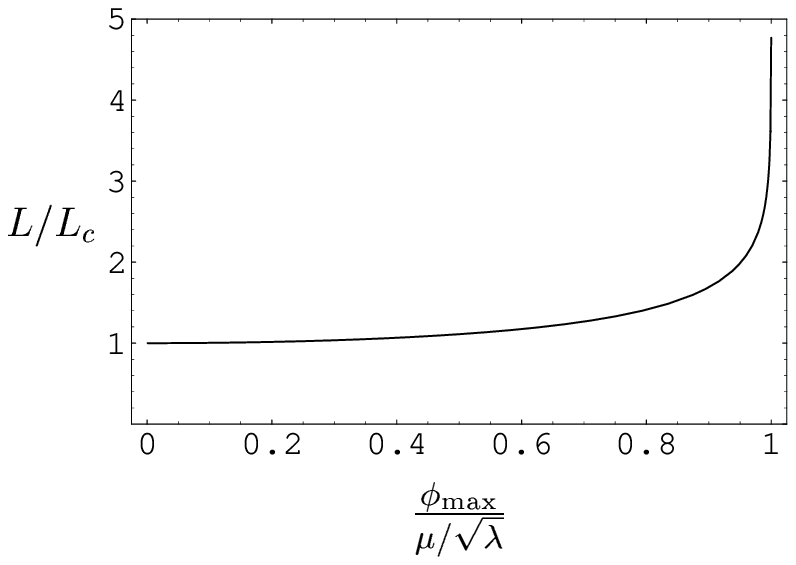}
   \centerline{\hspace{11mm}\bf (\ref{fig:kak}b)}}
   \caption{\small {\bf (\ref{fig:kak}a)}~A graph of the KAK-like solution;
      {\bf (\ref{fig:kak}b)}~The Orbifold length~$L$ vs. the
      amplitude~$\phi_{\rm max}$.}
   \label{fig:kak}
\end{figure}

We recall that given a scalar potential $U(\phi)$, the problem of finding
static solutions depending only on one spatial coordinate, is equivalent to
that of a non-relativistic particle~\cite{coleman} in the 1D potential $V=-U$.
The particle coordinate is $\phi$, and the ``time'' is $x_5$.
Considering the infinite dimension case (analogous to infinite
{\it time duration} in the mechanical analogy), the field must approach
two adjacent global minima at the boundaries (See Fig.~\ref{fig:orbit-1d}).
If $U$ has at least two global minima, there exist non-trivial solutions
interpolating between two adjacent global minima.
For example, the potential
\beq
   U(\phi)=\quart \lambda\le(\phi^2-\mu^2/\lambda\ri)^2,
   \label{eq:potential1}
\eeq
has two global minima:
$\phi_0=\pm \mu/\sqrt{\lambda}$, and thus the possible solutions are
the kink and the anti-kink:
\beq
   \phi(z)=\pm\frac{\mu}{\sqrt{\lambda}}
                  \tanh\frac{\mu z}{\sqrt{2}}.   \label{eq:kink}
\eeq
Proceeding to the more realistic case where the extra dimension is compact,
it is obvious that the kink solution is incompatible with plain periodical
boundary conditions.
A common solution to this problem is to impose $S^1/Z_2$ orbifold boundary
conditions, with the localizer odd under the $Z_2$.
Note that the orbifold $Z_2$ symmetry of the Lagrangian is reflected in
the fact that the solutions are either even or odd under the $Z_2$.
In the large-$L$ limit the localizer VEV
can be approximated by the kink-antikink (KAK) ansatz,
\beq
   \phi(z)=\phi_k(z)\ \phi_k(L-z) ; \quad H(z)={\rm const.}=v_H,
\eeq
which appears in numerous models~\cite{GP,GGH}.
This is understood by considering the localizer field $\phi$ with the
potential~(\ref{eq:potential1}).
The orbifold implications on the mechanical analogy are that we look for
a periodic motion with period $2L$ and with the analogue particle at the
origin (though not at rest) in the start and the end points of the period
(see Fig.~{\bf\ref{fig:kak}a}).
An explicit expression for this KAK-like solution is given by
\beq
   \z-\z_0=\pm\int_{\phi(\z_0)}^{\phi(\z)}\frac{d\phi}
   {\sqrt{2\le[U(\phi)-U(\phi_{\rm max})\ri]}}.  \label{eq:kink_int_orb}
\eeq
Unlike Eq.~(\ref{eq:kink}), this integral does not have a nice algebraic
form for our potential~\cite{GT}.
The relation between the orbifold size and the amplitude is given by
\beq
   L=\sqrt{2}\int_0^{\phi_{\rm max}}
   \frac{d\phi}{\sqrt{U(\phi)-U(\phi_{\rm max})}},
\eeq
with a numerical evaluation depicted in Fig.~{\bf \ref{fig:kak}b}.
For arbitrarily large-$L$ we can always find an appropriate motion whose
amplitude is arbitrarily close to the maximum of the potential energy
$\phi_{\rm max}=\mu/\sqrt{\lambda}$.
However, the period cannot be less than the small oscillation limit
$2\pi/\sqrt{-U''(0)}$.
Thus there is a critical orbifold size $L_c=\pi/\mu$ which is the
minimal one for a nontrivial solution.
For $L<L_c$ the only solution is the trivial one, $\phi(z) \equiv 0$.
The intuitive picture is that there is a tension between the ``natural''
VEV $\pm \mu/\sqrt{\lambda}$ and the orbifold boundary conditions which set
the odd field to zero at the fixed points.
If we compare the total energy of the two configurations - the KAK-like
vs. the identically-zero one, the total energy receives two kinds of
contributions, the first comes from the potential difference while the
second is the ``shear'' energy contributed from gradients in $z$.
Unlike the shear contribution, the potential contribution is proportional
to the bulk extent in which it resides, so we expect that when $L$
becomes small enough, the kinetic contribution takes over and the trivial
solution becomes more economical.
We also mention that the trivial configuration, $\phi(z) \equiv 0$, is
always a classical solution, possibly not a minimum of the action.
However, in the quantum theory such solution would tunnel to the true minimum
which is the nontrivial solution if it exists.

\subsection{Two Scalar Fields} \label{two}
In approaching the two scalar case, we find that a notation similar
to~\cite{rajaraman} can be useful.
In this notation, we rewrite the potential as
\beq
   U(\phi,h)=\quart \lambda\le(\phi^2-u^2\ri)^2+\half k^2 h^2
          +\quart\lambda_h h^4+\half g\ h^2 \le(\phi^2-u^2\ri),
   \label{eq:potential}
\eeq
where $\lambda,\lambda_h,g$ have dimension $-1$, $u$ has dimension $3/2$,
and $\lambda,\lambda_h,u^2>0$.
Comparing to Eq.(\ref{eq:two_pot}) we find the new parameters to be
\beq
   u^2\equiv \frac{\mu^2}{\lambda} \qquad \mbox{and} \qquad
   k^2\equiv \frac{g}{\lambda}\mu^2-\mu_h^2,  \label{eq:param}
\eeq
in the original notation.

\subsubsection{Infinite extra dimension}
\begin{figure}[t]
  \centering
  \makebox{\includegraphics[width=0.42\textwidth]{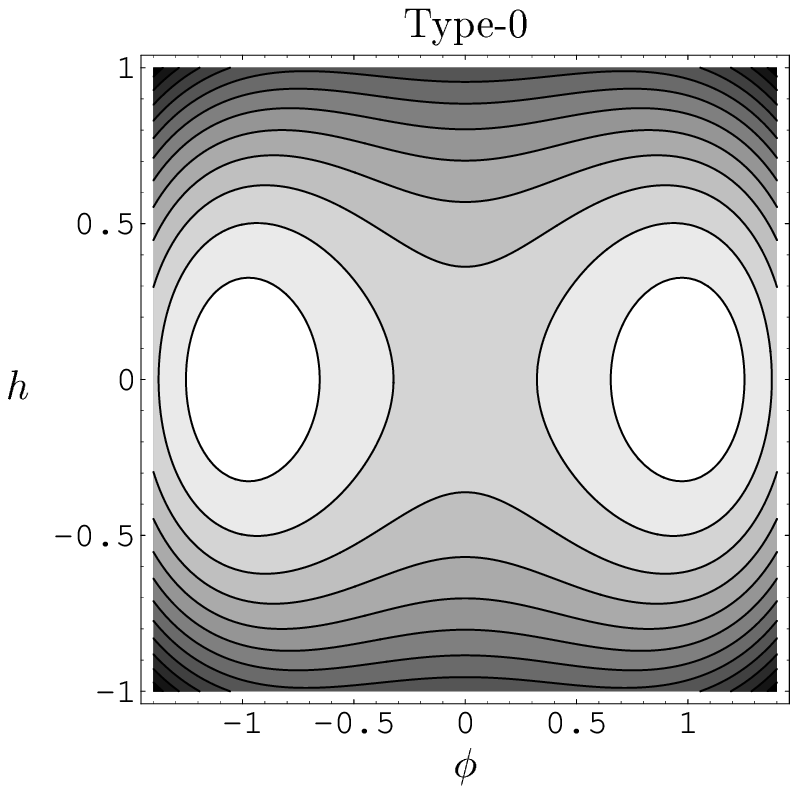}\qquad
  \quad
  \includegraphics[width=0.42\textwidth]{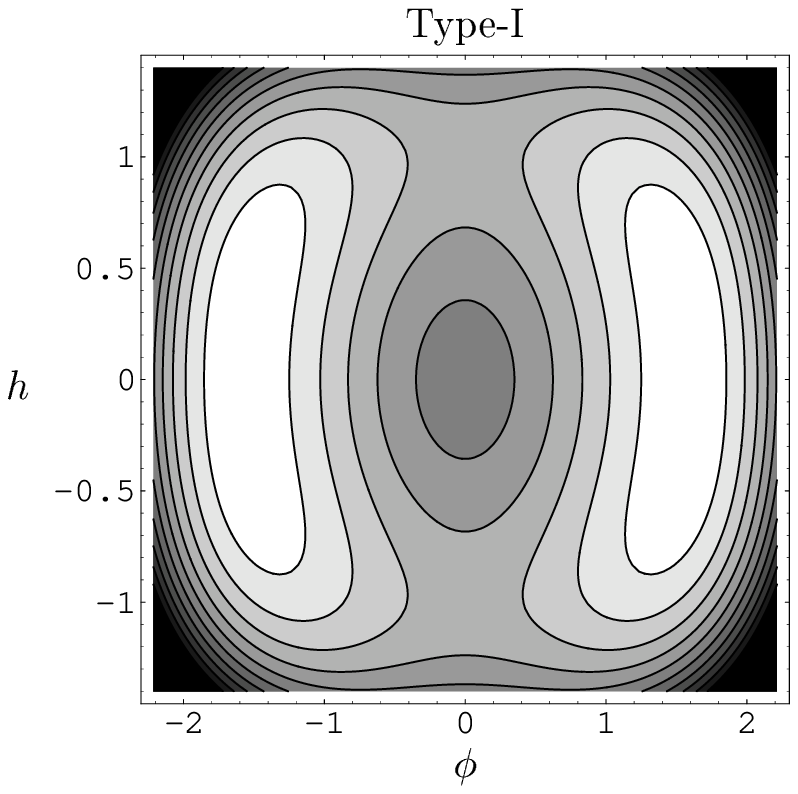}}\vspace{3mm}
  \makebox{\includegraphics[width=0.42\textwidth]{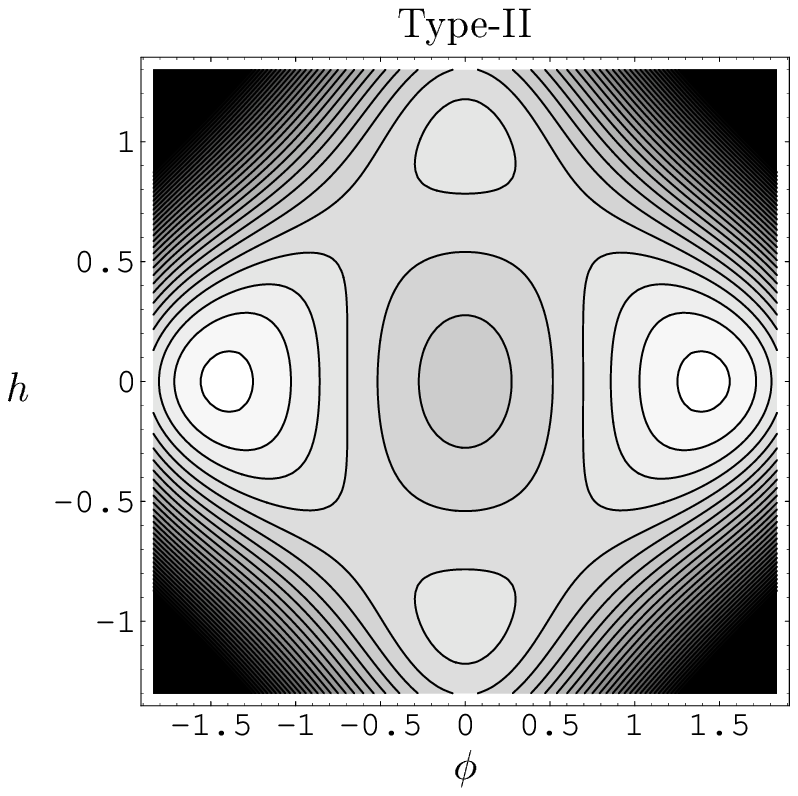}\qquad
  \quad \includegraphics[width=0.42\textwidth]{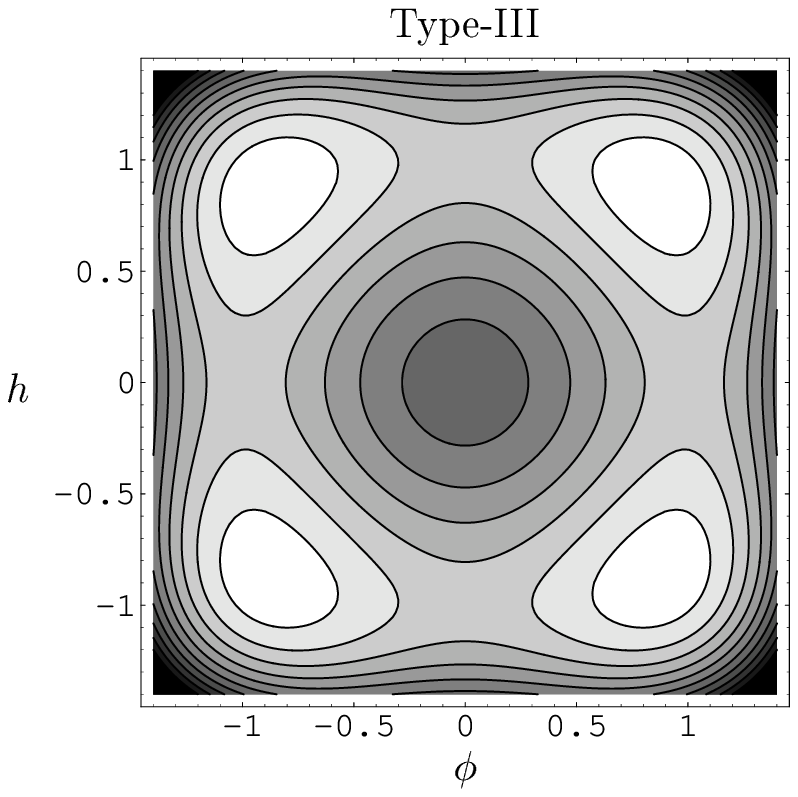}}
  \caption{\small Plots of isopotential lines for different potential
  types.
  The four different types are related to different regions in the
  parameter space, as explained in the text.
  The field values of $\phi$ and $h$ are divided by the
  factors $\mu/\sqrt{\lambda}$ and $\mu_h/\sqrt{\lambda_h}$
  accordingly.}  \label{fig:setups}
\end{figure}

The mechanical analogy for two scalar fields involves one particle in a
two dimensional potential.
A soliton is described by a zero-energy classical orbit starting
and ending at global maxima.
There are two types of such orbits:
non-topological orbits start and end in the same global minimum while
topological orbits connect two distinct global minima.
We focus on the latter type since the former one is non stable,
being in the same topological class as the trivial solution.
In order to classify the solitons, we should study the configuration of the
critical points in the potential.
We distinguish between four configurations of the critical points (see
Fig.~\ref{fig:setups}).
These four ``types'' of potentials are related to different regions in
the parameter space $(\mu_h,\mu,\lambda_{\phi},\lambda_h,g)$.
\begin{itemize}
\item {\bf Type 0:}
For this type of potentials, $\mu_h^2<0,\quad g\mu^2>\lambda\mu_h^2$.
This pattern has one maximum point at the origin and two
global minima at $(0,0),(\pm u,0)$.\\
\item {\bf Type-I:}
Another pattern occurs if
\beq
   0<\mu_h^2<\frac{g\mu^2}{\lambda} \qquad \mbox{and} \qquad
   \le(\frac{\mu_h}{\mu}\ri)^2 < \frac{\lambda_h}{g}.
\eeq
\item {\bf Type-II:}
An even richer pattern of critical points is achieved when
the last inequality is reversed,
\beq
   0<\mu_h^2<\frac{g\mu^2}{\lambda} \qquad \mbox{and} \qquad
   \le(\frac{\mu_h}{\mu}\ri)^2 > \frac{\lambda_h}{g}.
\eeq
\item {\bf Type-III:}
If the first inequality in Type-I is inverted, we get a pattern of four
global minima, one in each quadrant.
\end{itemize}
According to the mechanical analogy, type-0 does not yield a non trivial
solution for both fields.
Types~I-II are the ones for which we obtain exact solutions.
For type-III we conjecture qualitative characteristics of the
solution without proof.

Unlike the single field case, here one must guess an orbit in the $(\phi,h)$
plane.
If the guess is successful, an explicit solution as function of~$x_5$ can
be obtained.
However, even if our guess is successful,
the solution might be conditioned by constraints on the potential parameters.
The full scheme is explained in the appendix.
In Fig.~\ref{fig:raj}, an exact solution is depicted, along with its
corresponding mechanical orbit.
This solution is given explicitly by
\beq
   \phi(z)=u\tanh [k(z-z_0)] ; \qquad
   h(z)=\pm\sqrt{\frac{\lambda u^2-2k^2}{g}}\ {\rm sech}[k(z-z_0)],
   \label{eq:raj_sol0}
\eeq
with
\beq
   k=\sqrt{\frac{g}{\lambda}\mu^2-\mu_h^2}.  \label{eq:width}
\eeq
As can be seen in Fig.~\ref{fig:raj}, in this solution the localizer acquires
a kink-like profile and the Higgs VEV is bell-shaped, in accordance to the
perturbative approximation.
We note that the above solution has a constraint on the potential parameters,
which is
\beq
   \frac{\lambda_h}{\lambda} > \le(\frac{\mu_h}{\mu}\ri)^4 ; \qquad
   \le(\frac{\mu}{\mu_h}\ri)^2=\frac{2\lambda(g-\lambda_h)}
        {g^2-2g\lambda_h+\lambda\lambda_h}.  \label{eq:cond1}
\eeq
That is, the above nice algebraic solution is valid only with this constraint.
However, the mechanical analogy teaches us that similar solutions exist in the
neighborhood of Eq.(\ref{eq:cond1}), although they might not have
closed algebraic forms.
With a suitable choice of parameters, the above solution can serve as a
realization of the localized Higgs scenario.
In fact, as can be seen from Eq.(\ref{eq:width}), the $\mu_h/\mu$ hierarchy
makes sure that the above solution is tightly localized, provided that
$g/\lambda$ is not too small.
Another condition is the integral constraint Eq.(\ref{eq:phen}).
Putting the solution~(\ref{eq:raj_sol0}) into that constraint, we get
\beq
   \int_0^L dx_5\, h^2(x_5)=
   \frac{2}{k} \frac{\lambda u^2-2k^2}{g}=v^2_{\rm EW}
   \label{eq:realization},
\eeq
or
\beq
   v^2_{\rm EW}=\frac{2\mu^2\le(1-2\frac{g}{\lambda}\ri)+2\mu_h^2}
       {g\sqrt{\frac{g}{\lambda}\mu^2-\mu_h^2}}.
\eeq

\begin{figure}[t]
  \centering
  \makebox{\raisebox{0mm}{\includegraphics[width=0.45\textwidth]{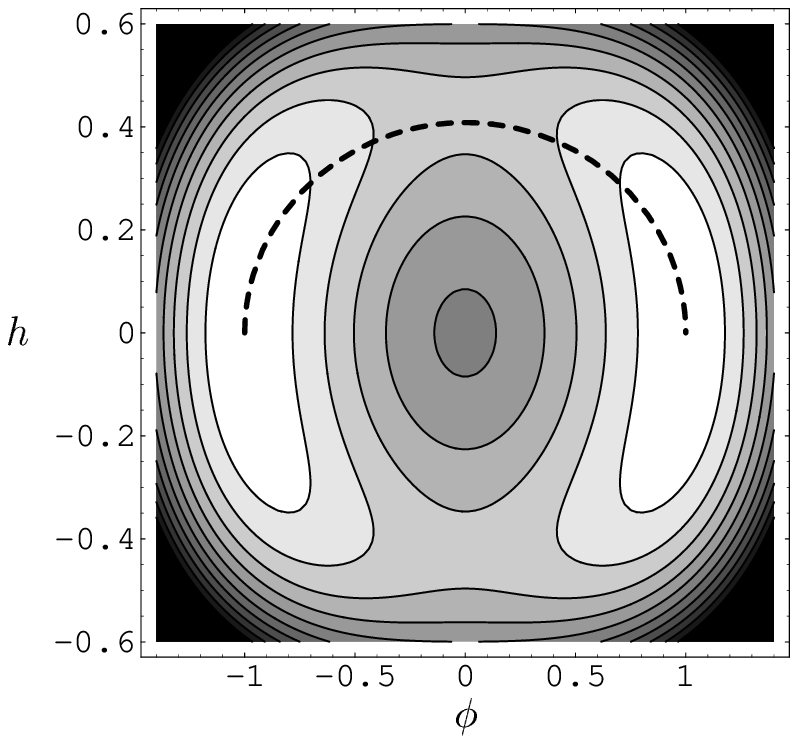}}\quad
  \includegraphics[width=0.5\textwidth]{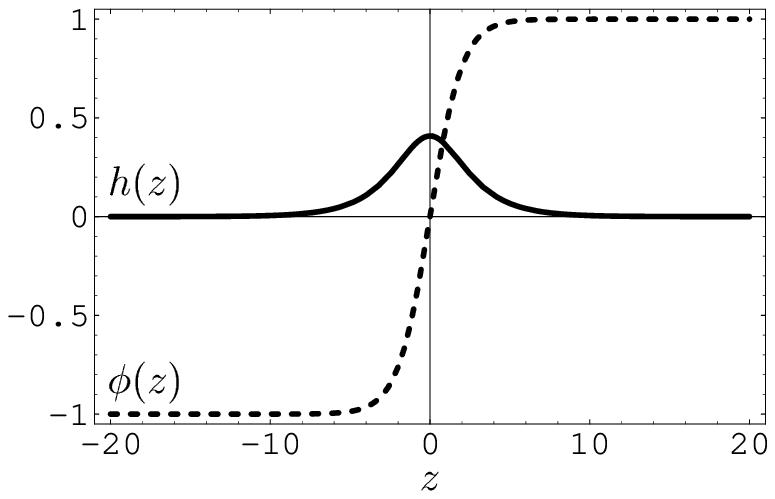}}
  {\bf \hspace*{6mm}(\ref{fig:raj}a)\hspace{68mm} (\ref{fig:raj}b)}
  \caption{\small An exact solution.
  Fig.~{\bf (\ref{fig:raj}a)} shows a half-elliptic orbit which yields the
     explicit solution in Fig.~{\bf (\ref{fig:raj}b)}.
  In Fig.~{\bf (\ref{fig:raj}b)} the solid (dashed) curve represents
  the Higgs (localizer) VEV.}
  \label{fig:raj}
\end{figure}

\begin{figure}[t]
  \centering
  \makebox{\includegraphics[width=0.4\textwidth]{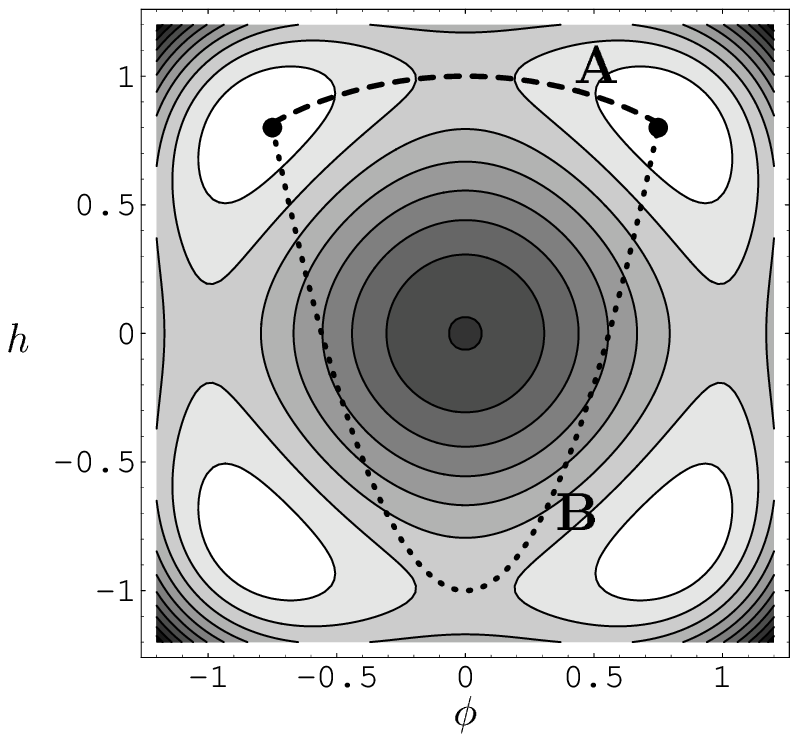}
       \qquad\includegraphics[width=0.5\textwidth]{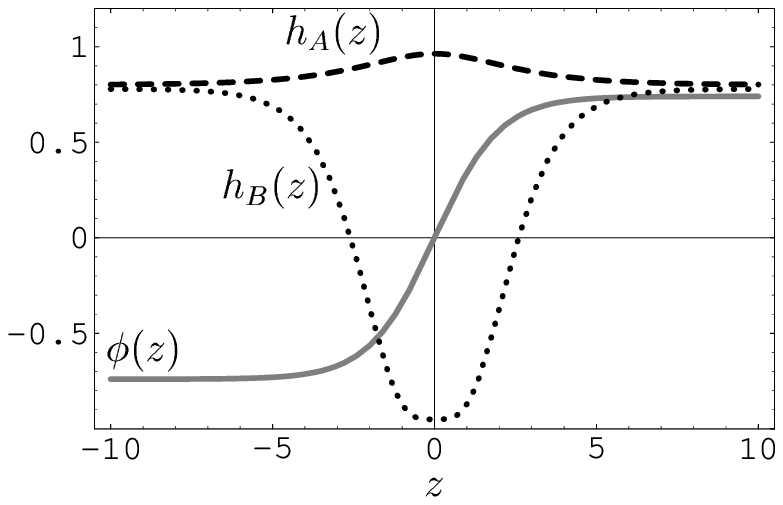}}
  {\bf \hspace*{3mm}(\ref{fig:type3}a)\hspace{71mm} (\ref{fig:type3}b)}
  \caption{\small Fig.~{\bf (\ref{fig:type3}a)} shows two conceivable
  orbits for a type-III potential.  Fig.~{\bf (\ref{fig:type3}b)} shows
  the corresponding VEVs.
  Note that these are not exact solutions but illustrations
  of conjectured solutions.}
  \label{fig:type3}
\end{figure}

In all the solutions above, $h(z)$ maintains the same sign.
This feature is expected to be violated in some of the solutions for
type-III potentials which have four global minima at
\beq
   (\phi,h)=
   \le(\pm \sqrt{\frac{g\mu_h^2-\lambda_h\mu^2}{g^2-\lambda\lambda_h}},
   \pm \sqrt{\frac{g\mu^2-\lambda\mu_h^2}{g^2-\lambda\lambda_h}} \ri).
\eeq
In this scenario the orbits which are relevant for our purposes are those
connecting between two adjacent maxima with the same value of $h$.
This is in order to match the orbifold boundary conditions.
There are two kinds of conceivable orbits with such a feature, which are
illustrated as ``A'' and ``B'' in Fig.~{\bf (\ref{fig:type3}a)}.
In solution ``A'' the Higgs is always positive (or negative),
realizing one of the ``adiabatic'' solutions depicted in
Fig.~\ref{fig:adiabatic},
while in solution~``B'' it becomes negative in the vicinity of $z=0$.
In both orbits the Localizer has a kink-like shape.
At this point we could not obtain an explicit form of neither solutions,
which may even not exist for the relevant region in the potential parameter
space.

Regarding the conjectured solution~``B'', two points are worth noting:
First, this solution cannot appear in the adiabatic solution since it
involves large gradients in $h(z)$, which is in contrast with the
``adiabatic'' assumption.
Second, unlike the neutral $h(z)$ in our simplified discussion,
the SM Higgs is charged under $SU(2)\times U(1)$, and thus the above picture
of ``sign alternating'' should be replaced by one where the SU(2) phase
rotates along the extra dimension.

\subsubsection{Orbifold}  \label{orbi2}
\begin{figure}[t]
  \centering
  \makebox{\includegraphics[width=0.47\textwidth]{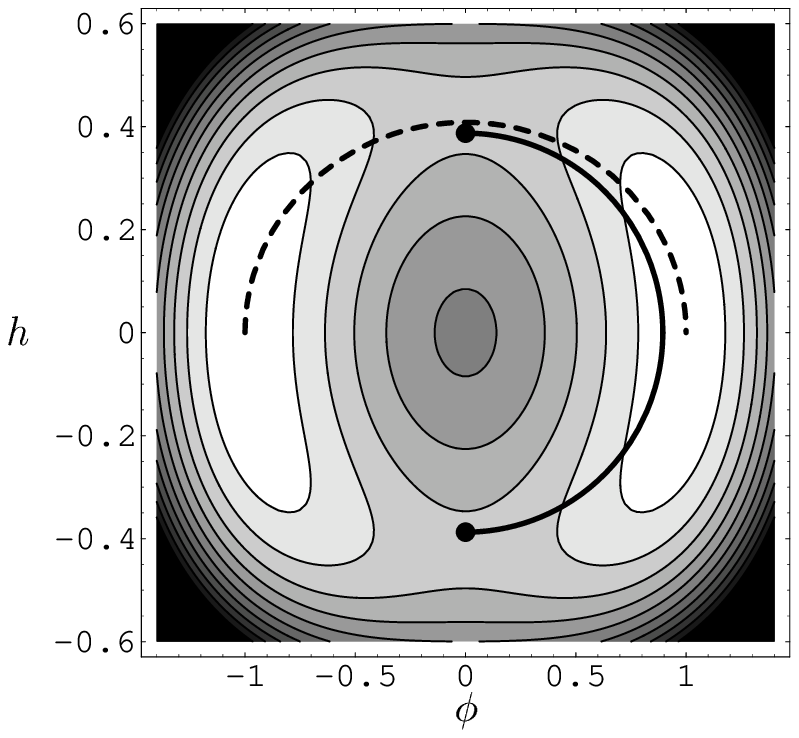}\quad
           \includegraphics{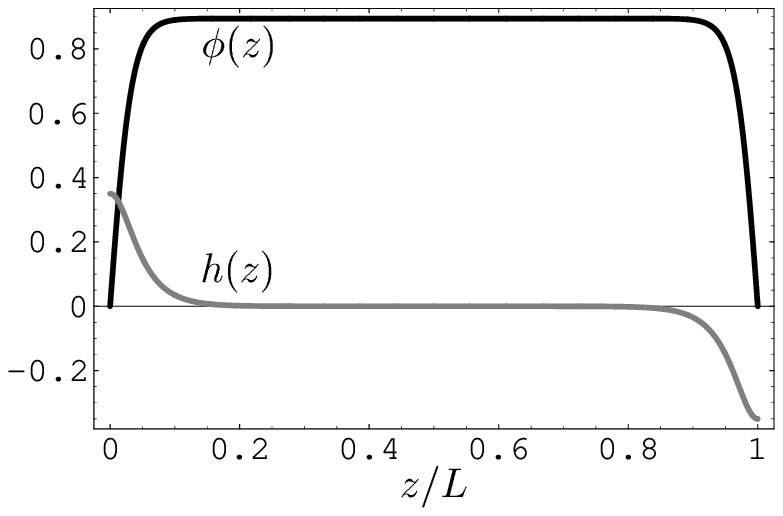}}
  {\bf \hspace*{9mm}(\ref{fig:orbi}a)\hspace{69mm} (\ref{fig:orbi}b)}
  \caption{\small An illustration of the conjectured solution for
     type-I potentials on the large $S^1/Z_2$ orbifold.
     Fig.~{\bf (\ref{fig:orbi}a)} shows the two orbits in large orbifold
     (solid line) and the $L\to \infty$ limit (dashed line).
     The orbifold fixed points are denoted by emphasized dots.
     Fig.~{\bf (\ref{fig:orbi}b)} illustrates the solution in the
     $L\to \infty$ limit.}
  \label{fig:orbi}
\end{figure}

\begin{figure}[t]
  \centering
  \includegraphics{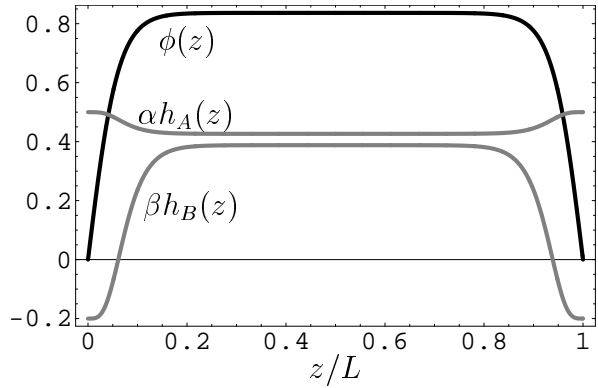}
  \caption{\small Conjectured solutions for type-III potentials on the
     $S^1/Z_2$ orbifold.  $A$ and $B$ denote the generalization of the
     orbits in figure~\ref{fig:type3}.  The scale factors $\alpha$ and $\beta$
     depend on the choice of the potential parameters.}
  \label{fig:type3-orbi}
\end{figure}

While in an infinite extra dimension we can obtain an explicit solution,
the orbifold case is much harder to solve.
Nevertheless some interesting conjectures can be made.
Again we seek oscillatory solutions with period $2L$.
We start by discussing the large-$L$ case.
In the case of infinite extra dimension, the analogue particle departs
from $(-\mu/\lambda,0)$ and arrives at $(+\mu/\lambda,0)$ in a zero energy
path (see Fig.~\ref{fig:orbi}, dashed line).
In a large but finite sized orbifold we expect closed (periodic) orbits
with negative energy rather than zero energy.
While we cannot prove nor verify the existence of such orbits,
below we suggest the main features of such solutions if they exist.
In an analogy to the single-field case, the particle starts at the fixed
point with $\phi(0)=0 ; h(0)\lesssim \sqrt{\alpha}u$,
with a non zero ``velocity'' ($\dot \phi\neq 0 ; \quad \dot h=0$).
The particle barely misses the maximum at $(+1,0)$, then it proceeds
to the opposite fixed point in a similar way.
The motion in the ``$\phi<0$'' plane is completely
dictated by the orbifold $Z_2$ symmetry (that is, in a perfect
reflection of the first half of the motion).
The period of the above motion amounts to the
complete circumference of the compact dimension (See Fig.~\ref{fig:orbi}).
The orbifold symmetry requirement is invariance
under reflection about the $h$-axis. In our case the symmetry of the
potential further implies that the orbit is also invariant under
reflection about the $\phi$-axis.  In the case of the $SU(2)\times
U(1)$ Higgs, the orbifold condition implies that the orbit is
invariant under reflection about the Higgs hyperplane.
We do not know if such an orbit (or a continuum of orbits)
exists, but we do know that if some orbit intersects the axes with right
angles $\le(\dot h(0)=0 , \quad \dot\phi(L/2)=0\ri)$ then it is periodic
because of symmetry considerations.
Here too, as in the case of one scalar field, there is a critical orbifold
size, under which the only solution is
\beq
   \phi(z)=0 ; \qquad h(z)=\pm\sqrt{\frac{g u^2-k^2}{\lambda_h}}=
   \pm\frac{\mu_h}{\sqrt{\lambda_h}},
   \label{eq:tiny_orb}
\eeq
since it always has a lower action than the trivial solution $(\phi,h)=(0,0)$.
This solution is incompatible with split fermion models.
For small but finite orbifold size, Eq.~(\ref{eq:tiny_orb}) cannot be
perturbed with small oscillations, since this point is not a minimum
of $-U(\phi,h)$.
Small oscillations can be found only about the origin, but since
the problem is equivalent to an anisotropic harmonic oscillator,
the only relevant solutions exist only when the ratio $\mu_h/\mu$ is
rational, which requires fine tuning of the mass parameters as well as of
the orbifold size.
More specifically, elliptic and circular orbits exist only if
$\mu=\mu_h$.
Another possibility regarding type-III potentials is depicted in
Fig.~\ref{fig:type3-orbi}.
This hypothetical solution is the orbifold version of the solutions ``A'' and
``B''.
In this case the Localizer is again KAK-like.

\section{A more generic scenario} \label{mildly}
As discussed before, the deviation of $v_H(z)$ from flatness is proportional
to $g$, the coupling of $\phi^2 H^{\dagger} H$.
Such deviations lead to Tree-Level FCNC, and therefore we can translate
the bounds from FCNC experimental data into a bound on $g$.
In order to do this, first we estimate the deviation using the
perturbativity condition Eq.(\ref{eq:region}).
In this limit it turns out that $f^h(z) \propto v_H(z)$ even if these
functions are not uniform.
This fact is demonstrated using the 5D point of view (see section~\ref{FCNC}).
We substitute the KK expansion,
\beq
   H(x,z)=v_H(z)+\sum_n h_n(x)f_n(z),
\eeq
into the 5D equation of motion.
Separation of variables for $f_n(z)$ yields
\beq
   f_n\dmu\partial^{\mu}h-hf''+(-\mu_h^2+g\phi^2)hf+\lambda(v_H+hf)^3=0.
\eeq
After linearization in $hf$ and separation, we obtain
\beq
   -f''+(-\mu_h^2+g\phi^2+3\lambda v_H^2)f=0,
\eeq
which, up to the scaling $v_H(z)\to v_H(z)/\sqrt{3}$ and upon substituting $f=v_H$,
is similar to the equation for the VEV,
\beq
   -v_H''+(-\mu_h^2+g\phi^2+\lambda v_H^2)v_H=0.
\eeq

This approximation holds as long as the localizer is not affected by the
Higgs VEV, namely when $gv_H^2(z) \ll \mu^2$,
and thus we expect that
\beq
   f(z) \propto v_H(z)+{\cal O}\le(\frac{g\mu_h^2}{\lambda_h\mu^2}\ri).
\eeq
This means that if the hierarchy $\mu_h^2/\mu^2$ is resolved, the
coupling $g$ does not have to be very small in order to suppress FCNC.
In particular, we are interested in tree-level processes where the Higgs KK
mediates flavor transitions.
For example, such an effective operator contributing to $K-\bar K$ mixing
is {\it e.g.:}
\beq
   \le(\frac{g\mu_h^2}{\lambda_h\mu^2}\ri)^2Y_{ij}^2
   \frac{s\bar ds\bar d}{m_{\rm KK}^2},
\eeq
where the Dirac structure is suppressed.
From experimental data of $K-\bar K$ mixing and CP violation in Kaon decay,
the suppression scale of a $s\bar ds\bar d/\Lambda^2$ term is
bounded by $\Lambda\gsim 10^4$~TeV~\cite{LP03}.
Following the rationale of some common flavor models (see
{\it e.g.}\cite{LNS}), we take
$Y_{sd} \simeq m_s/m_t \simeq (\sin\theta_c)^5$.
Furthermore, assuming $\lambda_h\sim 1$
one finds that there is no relevant bound on $g$.
Similar arguments hold for the $D^0$ and $B^0$ systems.
We conclude that configurations in which the Higgs VEV is neither uniform
nor confined to a brane do not impose further constraints
on the model parameters.

\section{Conclusions}
We discussed the implications of the Higgs coupling to the localizer
in split fermion models.
A scenario such as the Arkani-Hamed--Schmaltz model, where the Higgs VEV is
uniform, requires this coupling to be small, implying fine tuning
of the model parameters.
We found an exact classical solution for the case of an infinite extra
dimension, by applying a mechanical analogy~\cite{rajaraman}.
This solution, which is not fine tuned,
provides a realization of scenarios where the Higgs is confined to a
brane~\cite{KT}.
We also discussed qualitatively the more realistic case of a compact
extra dimension.
Furthermore, we showed that more generic configurations of the Higgs VEV
are phenomenologically viable.  The apparently dangerous Higgs mediated
FCNCs are suppressed already for \makebox{$g\sim {\cal O}(1)$}, since
they are proportional to $\mu_h^2/\mu^2$, which is
already assumed small in any split fermion model~\cite{KT}.

Many assumptions were made in constructing the simplified model of~\cite{AS}.
By now, most of these assumptions have been carefully studied and
showed to be, indeed, only simplifying ones.
The unrealistic infinite extra dimension is not needed when an orbifold is
used~\cite{GGH,KT,GP}.
The assumption that the coupling to the localizer and the bare mass term
can be diagonalized simultaneously, was relaxed in~\cite{twist}.
In this work we tested the assumption that the Higgs is flat.
We found that this assumption too is not a crucial one, providing further
reinforcement to the split fermion idea.

\acknowledgments
I am grateful to Yuval Grossman, Oleg Khasanov and Gilad Perez for their
extensive contribution to this work.
I am also indebted to Andrey Katz, Israel Klich, Yossi Nir, Martin Schmaltz,
Yael Shadmi and Tomer Volansky for their help throughout this work and for
many fruitful discussions.

\appendix

\section{Exact Solutions}   \label{raj-orbits}
In this appendix we give some details of obtaining the exact
solution~(\ref{eq:raj_sol0}) and other solutions.
Here we follow a line similar to Rajaraman~\cite{rajaraman}.

\subsection{Orbits in the Mechanical Analogy}
A first integration of the equations of motion~(\ref{eq:eom2}) yields
two coupled ordinary differential equations,
 \beq
   \half \phi'^2=\int\pd{U(\phi,h)}{\phi}\ g\phi +A ; \qquad
   \half h'^2=\int\pd{U(\phi,h)}{h}\ g\ h +B, \label{eq:eom2d}
\eeq
compared with the one scalar case where there is one equation
only.
Here a primed field denotes its derivative with respect to $z$,
and $A,B$ are integration constants.

A solution may be obtained as follows.
First, we guess an equation for the mechanical orbit:
\beq
   g(\phi,h)=0. \label{eq:orbit}
\eeq
Differentiating both sides of (\ref{eq:orbit}) with respect to $z$ and
squaring, yields
\beq
   \le(\frac{\partial g}{\partial \phi}\ri)^2\phi'^2
   =\le(\frac{\partial g}{\partial h}\ri)^2 h'^2 .
\eeq
Inserting (\ref{eq:eom2d}), we obtain
\beq
   \le(\pd{g}{\phi}\ri)^2\le(\int_{\rm orbit}\pd{U(\phi,h)}{\phi}\
       g\phi \ri) =
   \le(\pd{g}{h}\ri)^2\le(\int_{\rm orbit}\pd{U(\phi,h)}{h}\
       g\ h \ri),  \label{eq:raj}
\eeq
where the integrals are evaluated along the orbit (\ref{eq:orbit}).
Eq.~(\ref{eq:raj}) imposes relations among the parameters in (\ref{eq:orbit})
and those in the potential.
Thus in general we must not expect that only the orbit parameters are
constrained, while those of the potential remain intact, unless
our guess of Eq.~(\ref{eq:orbit}) is exceptionally successful.

The most obvious orbit connecting the two vacua is the straight line
from $(\phi,h)=(-u,0)$ to $(+u,0)$.
A somewhat more complicated orbit could be the following:
Consider the one parameter family of canonical ellipses which go
through $(0,\pm u)$
\beq
   g(\phi,h)=h^2+\alpha(\phi^2-u^2)=0,  \label{eq:orbit1}
\eeq
where the orbit parameterization is such that it starts ($z\to-\infty$)
at $(\phi,h)=(-u,0)$ and ends ($z\to +\infty$) at $(+u,0)$.
By differentiating we find that
\beq
   \le(\pd{g}{\phi}\ri)^2=4\alpha^2\phi^2=4\alpha\le(\alpha u^2-h^2\ri)
   ; \qquad  \le(\pd{g}{h}\ri)^2=4h^2,
\eeq
and that
\beq
   dh^2=-\alpha\ d\phi^2.
\eeq
The last relation is used for calculating the integrals
\beqa
   \int\le(\pd{U}{\phi}\ri)\ d\phi &=&
   \half\int\le[\lambda(\phi^2-u^2)+g\ h^2\ri]\ d\phi^2 =
   \frac{\lambda-\alpha g}{2\alpha^2}\int h^2\ dh^2 \nonumber \\
   &=&  \frac{\lambda-\alpha g}{4\alpha^2}h^4
\eeqa
and
\beqa
   \int\le(\pd{U}{h}\ri)\ d\ h &=&
   \half\int (\lambda_h h^2+k^2-g u^2+g\phi^2)\ dh^2 =
   \half\int \le(\frac{\alpha\lambda_h-g}{\alpha}h^2+k^2\ri)\ dh^2
         \nonumber \\ &=&
   \frac{h^2}{4}\le(\frac{\alpha\lambda_h-g}{\alpha}h^2+2k^2\ri).
\eeqa
Substituting in (\ref{eq:raj}) we get
\beq
   (\lambda-\alpha g)\le(\alpha u^2-h^2\ri)h^4 =
   \le[2k^2\alpha+(\alpha\lambda_h-g)h^2\ri]h^4,
\eeq
or
\beq
   \alpha\le(\lambda u^2-\alpha g u^2-2k^2\ri)+
   \le[\lambda+g-\alpha(g+\lambda_h)\ri]h^2=0.
\eeq
The above must vanish identically, giving
\beq
   \alpha=\frac{\lambda u^2-2k^2}{g u^2} ; \qquad
   \lambda_h=\frac{g\le(g u^2-2k^2\ri)}{\lambda u^2-2k^2}.
   \label{eq:cond_raj}
\eeq
Since there is only one free parameter ($\alpha$), we have one equation
too many.
While the first condition gives us the desired orbit, the second relation
rather constrains the potential.
At this point we may wonder whether this illness could be remedied by
replacing the orbit equation (\ref{eq:orbit1}) with a {\it two} parameter
family such
as~\footnote{Note that both families of curves include also the parabolic
orbit $h+\alpha(\phi^2-u^2)=0$, which yields a solution with similar
characteristics as the elliptic orbit discussed above.}
\beq
   h^2+\alpha\le(\phi^2-u^2\ri)+\beta\le(\phi^2-u^2\ri)^2=0
   \quad {\rm or} \quad
   \alpha h^2+\beta h+\le(\phi^2-u^2\ri)=0.
\eeq
As we found out, the answer is negative.
Substituting the above two-parameter orbit in (\ref{eq:raj}) we do find,
in addition to the two Rajaraman solutions, solutions with
$\alpha,\beta \neq 0$.
However, the constraints on the potential parameters are not relaxed.
For example the orbit
\beq
   \alpha=\beta=\frac{\lambda u^2-2k^2}{g u^2},
\eeq
is a viable orbit only if the following two constraints are met:
\beq
   g u^2=2\le(\lambda u^2+k^2\ri); \qquad \lambda_h=\frac{8g}{3}.
\eeq

\subsection{Explicit Solutions}   \label{raj-sol}
With a legitimate one-particle orbit at hand we can finally decouple the
equations of motion (\ref{eq:eom2}) by substituting the orbit.
For the straight line ($h=0$) we have
\beq
   \phi''=\lambda\phi^3-\lambda u^2\phi+g\phi h^2=
   \lambda \phi^3-\lambda u^2\phi,
\eeq
which is solved by
\beq
   \phi(z)=u\tanh \le[\sqrt{\frac{\lambda}{2}}u(z-z_0)\ri] ; \qquad   h(z)=0.
   \label{eq:straight}
\eeq
For the more interesting orbit (\ref{eq:orbit1}) we have
\beq
   \phi''=\lambda\phi^3-\lambda u^2\phi+g\phi h^2=
          \frac{2k^2}{u^2}\phi^3-2k^2\phi,
\eeq
whose solution is
\beq
   \phi(z)=u\tanh [k(z-z_0)] ; \qquad
   h(z)=\pm\sqrt{\frac{\lambda u^2-2k^2}{g}}\ {\rm sech}[k(z-z_0)].
\eeq
There is no apparent reason that one of the above solutions has
the globally minimal action.
In absence of a uniqueness theorem for such nonlinear equations,
we can only rule out a solution if we find another solution with
smaller action.
Considering the above solutions, we find that their actions are
\beq
   S_{\rm straight}=-\frac{2}{3}\sqrt{2\lambda} ; \qquad
   S_{\rm ellipse}=-\frac{2}{3}\frac{k}{g u}
      \le(\lambda+2g-\frac{2k^2}{u^2}\ri).
\eeq
Provided the conditions (\ref{eq:cond1}) are met,
sometimes the straight line has a smaller action than the elliptic orbit
and sometimes it is vice versa, depending on the potential parameters.
Moreover, these two trajectories might be not minima but maxima or saddle
points.



\begin{thebibliography}{99}

\bibitem{AS}
N.~Arkani-Hamed and M.~Schmaltz,
Phys.\ Rev.\ D {\bf 61}, 033005 (2000) [hep-ph/9903417].

\bibitem{MS}
E.~A.~Mirabelli and M.~Schmaltz,
Phys.\ Rev.\ D {\bf 61}, 113011 (2000)
[hep-ph/9912265].

\bibitem{GP}
Y.~Grossman and G.~Perez,
Phys.\ Rev.\ D {\bf 67}, 015011 (2003)
[hep-ph/0210053].

\bibitem{ASmodels}
J.~L.~Crooks, J.~O.~Dunn and P.~H.~Frampton,
Astrophys.\ J.\ {\bf 546}, L1 (2001)
[astro-ph/0002089];
G.~C.~Branco, A.~de Gouvea and M.~N.~Rebelo,
Phys.\ Lett.\ B {\bf 506}, 115 (2001)
[hep-ph/0012289];
W.~F.~Chang and J.~N.~Ng,
P.~Q.~Hung and M.~Seco,
Nucl.\ Phys.\ B {\bf 653}, 123 (2003)
[hep-ph/0111013];
Y.~Uehara,
JHEP {\bf 0112}, 034 (2001)
[hep-ph/0107297];
G.~Barenboim, G.~C.~Branco, A.~de Gouvea and M.~N.~Rebelo,
Phys.\ Rev.\ D {\bf 64}, 073005 (2001)
[hep-ph/0104312];
W.~F.~Chang and J.~N.~Ng, JHEP {\bf 0212}, 077 (2002)
[hep-ph/0210414];
H.~V.~Klapdor-Kleingrothaus and U.~Sarkar,
Phys.\ Lett.\ B {\bf 541}, 332 (2002)
[hep-ph/0201226];
B.~Lillie and J.~L.~Hewett,
Phys.\ Rev.\ D {\bf 68}, 116002 (2003)
[hep-ph/0306193];
P.~Q.~Hung,
Phys.\ Rev.\ D {\bf 67}, 095011 (2003)
[hep-ph/0210131];
S.~Khalil and R.~Mohapatra,
Nucl.\ Phys.\ B {\bf 695}, 313 (2004)
[hep-ph/0402225];
P.~Q.~Hung, M.~Seco and A.~Soddu,
Nucl.\ Phys.\ B {\bf 692}, 83 (2004)
[hep-ph/0311198];
J.~M.~Frere, G.~Moreau and E.~Nezri,
Phys.\ Rev.\ D {\bf 69}, 033003 (2004)
[hep-ph/0309218];
H.~Abe, K.~Choi, K.~S.~Jeong and K.~i.~Okumura
  [hep-ph/0409196];
S.~Khalil and R.~Mohapatra,
  Nucl.\ Phys.\ B {\bf 695}, 313 (2004)
  [hep-ph/0402225];
H.~D.~Kim, S.~Raby and L.~Schradin
  [hep-ph/0411328];
Y.~Nagatani and G.~Perez,
JHEP {\bf 0502}, 068 (2005)
[hep-ph/0401070];
R.~Harnik, G.~Perez, M.~D.~Schwartz and Y.~Shirman,
  JHEP {\bf 0503}, 068 (2005)
  [hep-ph/0411132];
M.~Gogberashvili, P.~Midodashvili and D.~Singleton
  [hep-th/0505030];
E.~O.~Iltan
  [hep-ph/0504013,
  hep-ph/0503001];
G.~A.~Palma
  [hep-th/0505170];


\bibitem{KT}
D.~E.~Kaplan and T.~M.~Tait,
JHEP {\bf 0111}, 051 (2001)
[hep-ph/0110126].

\bibitem{GaussLocH}
N.~Maru,
Phys.\ Lett.\ B {\bf 522}, 117 (2001)
[hep-ph/0108002];
M.~Kakizaki and M.~Yamaguchi,
Int.\ J.\ Mod.\ Phys.\ A {\bf 19}, 1715 (2004)
[hep-ph/0110266];
M.~Maru, N.~Sakai, Y.~Sakamura and R.~Sugisaka,
Nucl.\ Phys.\ B {\bf 616}, 47 (2001)
[hep-th/0107204];
M.~Kakizaki and M.~Yamaguchi,
Prog.\ Theor.\ Phys.\ {\bf 107}, 433 (2002)
[hep-ph/0104103];
N.~Haba and N.~Maru:
Phys.\ Rev.\ D {\bf 66}, 055005 (2002)
[hep-ph/0204069],
Mod.\ Phys.\ Lett.\ A {\bf 17}, 2341 (2002)
[hep-ph/0202196],
Phys.\ Lett.\ B {\bf 532}, 93 (2002)
[hep-ph/0201216];
N.~Haba, N.~Maru and N.~Nakamura,
Phys.\ Lett.\ B {\bf 557}, 240 (2003)
[hep-ph/0209009].


\bibitem{LocH}
D.~E.~Kaplan and T.~M.~Tait,
JHEP {\bf 0006}, 020 (2000)
[hep-ph/0004200];
F.~Del Aguila and J.~Santiago,
JHEP {\bf 0203}, 010 (2002)
[hep-ph/0111047];
A.~Hebecker and J.~March-Russell,
Phys.\ Lett.\ B {\bf 541}, 338 (2002)
[hep-ph/0205143];
G.~Perez,
Phys.\ Rev.\ D {\bf 67}, 013009 (2003)
[hep-ph/0208102];
B.~Lillie,
  JHEP {\bf 0312}, 030 (2003)
  [hep-ph/0308091];
G.~Bhattacharyya, H.~V.~Klapdor-Kleingrothaus, H.~Pas and A.~Pilaftsis
  [hep-ph/0402071];
G.~Perez and T.~Volansky
  [hep-ph/0505222].

\bibitem{PDG}
J.~Hewett and J.~March-Russell, in K.~Hagiwara {\it et al.},
Phys.\ Rev.\ D {\bf 66}, 010001 (2002).

\bibitem{GGH}
H.~Georgi, A.~K.~Grant and G.~Hailu,
Phys.\ Rev.\ D {\bf 63}, 064027 (2001)
[hep-ph/0007350].

\bibitem{AGS}
N.~Arkani-Hamed, Y.~Grossman and M.~Schmaltz,
Phys.\ Rev.\ D {\bf 61}, 115004 (2000) [hep-ph/9909411];
T.~Han, G.~D.~Kribs and B.~McElrath,
Phys.\ Rev.\ Lett.\  {\bf 90}, 031601 (2003)
[hep-ph/0207003];
W.~F.~Chang and J.~N.~Ng,
  JHEP {\bf 0212}, 077 (2002)
  [hep-ph/0210414];
W.~F.~Chang, I.~L.~Ho and J.~N.~Ng,
Phys.\ Rev.\ D {\bf 66}, 076004 (2002)
[hep-ph/0203212];
S.~Nussinov and R.~Shrock,
Phys.\ Lett.\ B {\bf 526}, 137 (2002)
[hep-ph/0101340];
Phys.\ Rev.\ Lett.\ {\bf 88}, 171601 (2002)
[hep-ph/0112337];
D.~J.~Chung and T.~Dent,
Phys.\ Rev.\ D {\bf 66}, 023501 (2002)
[hep-ph/0112360];
T.~G.~Rizzo,
Phys.\ Rev.\ D {\bf 64}, 015003 (2001)
[hep-ph/0101278];
A.~Masiero, M.~Peloso, L.~Sorbo and R.~Tabbash,
Phys.\ Rev.\ D {\bf 62}, 063515 (2000)
[hep-ph/0003312];
A.~Delgado, A.~Pomarol and M.~Quiros,
JHEP {\bf 0001}, 030 (2000)
[hep-ph/9911252].

\bibitem{DS}
G.~R.~Dvali and M.~A.~Shifman,
Phys.\ Lett.\ B {\bf 475}, 295 (2000)
[hep-ph/0001072].

\bibitem{twist}
Y.~Grossman, R.~Harnik, G.~Perez, M.~D.~Schwartz and Z.~Surujon,
Phys.\ Rev.\ D {\bf 71}, 056007 (2005)
[hep-ph/0407260].

\bibitem{anom}
  N.~Arkani-Hamed, A.~G.~Cohen and H.~Georgi,
  Phys.\ Lett.\ B {\bf 516}, 395 (2001)
  [hep-th/0103135];
  C.~A.~Scrucca, M.~Serone, L.~Silvestrini and F.~Zwirner,
  Phys.\ Lett.\ B {\bf 525}, 169 (2002)
  [hep-th/0110073];
  S.~Groot Nibbelink, H.~P.~Nilles and M.~Olechowski,
  Nucl.\ Phys.\ B {\bf 640}, 171 (2002)
  [hep-th/0205012];
  G.~von Gersdorff and M.~Quiros,
  Phys.\ Rev.\ D {\bf 68}, 105002 (2003)
  [hep-th/0305024].

\bibitem{coleman}
S.~Coleman,
{\it Aspects of Symmetry}, Cambridge University Press 1985.

\bibitem{rajaraman}
R.~Rajaraman,
{\it Solitons and Instantons},
Amsterdam: North Holland, 1982.

\bibitem{GT}
B.~Grzadkowski and M.~Toharia,
Nucl.\ Phys.\ B {\bf 686}, 165 (2004)
[hep-ph/0401108].

\bibitem{LP03}
Y.~Grossman,
Int.\ J.\ Mod.\ Phys.\ A {\bf 19}, 907 (2004)
[hep-ph/0310229].

\bibitem{LNS}
  M.~Leurer, Y.~Nir and N.~Seiberg,
  Nucl.\ Phys.\ B {\bf 420}, 468 (1994)
  [hep-ph/9310320].

\end{thebibliography}
\end{document}